\documentclass[12pt, a4, preprint]{aastex}

\slugcomment{Accepted for publication in ApJL, 12 August 2014}

\shorttitle{The albedo-color diversity of TNOs}
\shortauthors{Pedro Lacerda and the TNOs Are Cool Team}

\title{The albedo-color diversity of transneptunian objects}

\author{Pedro Lacerda$^{1}$, Sonia Fornasier$^2$, Emmanuel
  Lellouch$^2$, Csaba Kiss$^3$, Esa Vilenius$^4$, Pablo
  Santos-Sanz$^5$, Miriam Rengel$^1$, Thomas M\"uller$^4$, John
  Stansberry$^6$, Ren\'e Duffard$^5$, Audrey Delsanti$^{7,2}$, \&
Aur\'elie Guilbert-Lepoutre$^8$}

  \affil{1) Max-Planck-Institut f\"ur Sonnensystemforschung,
  Justus-von-Liebig-Weg 3, 37077 G\"ottingen, Germany}
  \affil{2) LESIA-Observatoire de Paris, CNRS, UPMC, Universit\'e
  Paris-Diderot, 5 place Jules Janssen, 92195 Meudon, France}
  \affil{3) Konkoly Observatory, MTA CSFK, 1121 Budapest, Konkoly Th. M.
  \'ut 15-17, Hungary}
  \affil{4) Max-Planck-Institut f\"ur extraterrestrische Physik,
  Postfach 1312, Giessenbachstrasse, 85741 Garching, Germany}
  \affil{5) Instituto de Astrof\'isica de Andaluc\'ia (IAA-CSIC) Glorieta
  de la Astronom\'ia, s/n 18008 Granada, Spain}
  \affil{6) Space Telescope Science Institute, 3700 San Martin Drive,
  Baltimore MD 21218, USA}
  \affil{7) Aix Marseille Universit\'e, CNRS, LAM (Laboratoire
  d'Astrophysique de Marseille) UMR 7326, 13388, Marseille, France}
  \affil{8) European Space Agency / ESTEC, Keplerlaan 1, 2201 AZ
  Noordwijk, Netherlands}

\begin{document}

\begin{abstract}

We analyze albedo data obtained using the Herschel Space Observatory
that reveal the existence of two distinct types of surface among
midsized transneptunian objects. A color-albedo diagram shows two
large clusters of objects, one redder and higher albedo and another
darker and more neutrally colored. Crucially, all objects in our
sample located in dynamically stable orbits within the classical
Kuiper belt region and beyond are confined to the bright-red group,
implying a compositional link. Those objects are believed to have
formed further from the Sun than the dark-neutral bodies.  This
color-albedo separation is evidence for a compositional discontinuity
in the young solar system.

\end{abstract}


\section{Introduction}

\begin{figure}
  \epsscale{0.7}
  \plotone{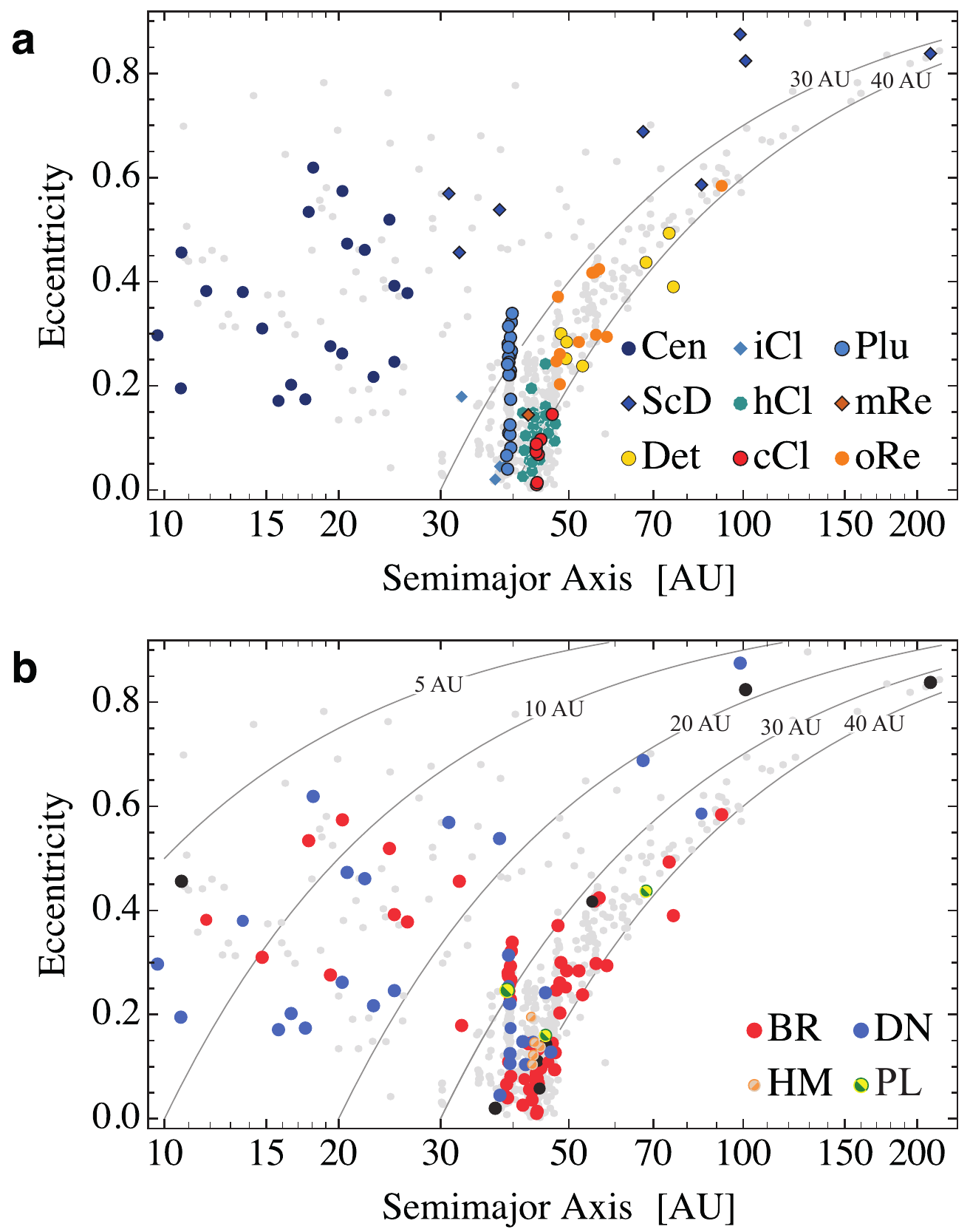}

  \caption{a) Orbital distribution (semimajor axis vs.\ eccentricity)
    of our sample (Cen=Centaurs, iCl=Inner Classicals, Plu=Plutinos,
    ScD=Scattered Disc, hCl=Hot Classicals, mRe=Middle Resonants,
    Det=Detached, cCl=Cold Classicals, oRe=Outer Resonants).  Light
    gray points mark TNOs not observed by Herschel.  Curves of
    constant perihelion are plotted in solid gray and labelled.  b)
    Same as a) but color-coded by surface type (BR=Bright Red, DN=Dark
  Neutral, HM=Haumea-type, PL=Dwarf Planets, black points have large
uncertainties and ambiguous surface type.)  }

  \label{Fig.ae}
\end{figure}

In the current paradigm, developed largely to explain the orbits of
transneptunian objects (TNOs), the dynamical architecture of the solar
system is thought to have evolved considerably since formation,
particularly in the first billion years
\citep{2005Natur.435..459Tsiganis}. A violent planetary instability
involving Jupiter and Saturn is hypothesized to have caused Uranus and
Neptune to migrate from their formation region (within 15 AU) to their
current orbits. This event led to the stochastic dispersal of the
planetesimal disk and resulted in bodies formed at various distances
from the Sun being stored together in the transneptunian space
\citep{2008Icar..196..258Levison}.  This dynamical restructuring may
explain the broad diversity seen in the properties of TNOs, but could
also obscure links with birth location that carry information about
the properties of the protoplanetary disk.  One exception is the Cold
Classical population (Figure \ref{Fig.ae}), which stands out as
possessing a number of properties that suggest a unique origin and
evolution \citep{2008ssbn.book...43Gladman}. Cold Classicals have low
inclination orbits in the region known as the classical Kuiper belt, a
donut-shaped structure located between the 3:2 and the 2:1 Neptunian
mean-motion resonances (MMRs) at 39 and 47 AU.  Dynamically, these
objects show no signs of past interactions with Neptune: they are
decoupled from the ice giant and are stable on Gyr timescales
\citep{2011ApJ...738...13Batygin}.  Physically, when compared to other
TNOs, Cold Classicals possess redder surfaces
\citep{2000Natur.407..979Tegler}, smaller sizes
\citep{2001AJ....121.1730Levison}, and a much larger abundance of
binaries \citep{2006AJ....131.1142Stephens}, including weakly bound
pairs that would have been disrupted by encounters with planets
\citep{2010ApJ...722L.204Parker}.  Attempts to explain the origin of
this population nearer the Sun and their transport out to the
classical Kuiper belt following the planetary instability have been
unsuccessful \citep[][but see
\cite{2014Icar..232...81Morbidelli}]{2003Natur.426..419Levison,
2008Icar..196..258Levison}.  Consequently, Cold Classicals are most
simply understood as survivors of an original population that formed
in-situ \citep{2001AJ....121.1730Levison, 2011ApJ...738...13Batygin}
and, as such, are representative of the properties of bodies that
originally formed beyond Neptune.  In this Letter, we analyze albedo
data for 109 TNOs and Centaurs obtained using the Herschel Space
Observatory and find that Cold Classical TNOs are not as unique in
terms of their surface properties as previously believed. Indeed, all
sampled TNOs in dynamical classes thought to originate beyond Neptune
display similar color/albedo properties.

\section{Herschel ``TNOs Are Cool'' Survey and Sample}

Herschel was the first large aperture (3.5 meter) space telescope to
operate in the far-infrared and submillimeter and it offered a unique
chance to measure the thermal radiation from the cool TNOs
(equilibrium temperatures $\sim$40 K).  Starting in 2009, we conducted
the ``TNOs Are Cool'' survey of the outer solar system using Herschel
to measure albedos and sizes for 130 TNOs and Centaurs
\citep{2009EM&P..105..209Mueller}, tripling the size of the existing
sample.  Thermal observations obtained earlier using Spitzer Space
Telescope \citep{2008prpl.conf..161Stansberry} were used to complement
the Herschel data, and a handful of objects in our sample have
independent, highly accurate diameter (and albedo) estimates from
stellar occultations. The thermal data were combined with existing
optical data to derive albedos and diameters using the radiometric
method or more detailed thermophysical models when justified.  Details
of the survey and modeling of albedos and diameters have been
published in the ``TNOs Are Cool'' series of papers \citep[e.g.,][and Kiss
et al.\ in prep.]{2010A&A...518L.146Mueller,
  2010A&A...518L.147Lellouch, 2010A&A...518L.148Lim,
  2012A&A...541A..92Santos-Sanz, 2012A&A...541A..93Mommert,
  2012A&A...541A..94Vilenius, 2012A&A...541L...6Pal,
  2013A&A...555A..15Fornasier, 2013A&A...557A..60Lellouch,
  2013ExA...tmp...36Kiss, 2014A&A...564A..35Vilenius}. 

Our sample covers the different dynamical classes to enable
comparative studies.  We adopt a widely accepted dynamical
classification scheme \citep{2008ssbn.book...43Gladman} that groups
TNOs into Resonant, Scattered, Detached, Classical (Figure
\ref{Fig.ae}; Table\ \ref{Table.One}).  We split the Classicals into
Hot and Cold at orbital inclination $i=5$ deg, and the Resonants into
Inner (objects in the 3:2 MMR and in resonances nearer the Sun),
Middle (located between the 3:2 MMR and the 2:1 MMR), and Outer (2:1
MMR and beyond). In our data, the Inner Resonants are represented by
21 Plutinos (3:2 MMR), the Middle Resonants consist of a single TNO
(5:3 MMR), and the Outer Resonants include 4 TNOs in the 2:1 MMR, 1 in
the 9:4, 4 in the 5:2, 1 in the 8:3 and 1 in the 11:2.  Our sample
also includes 4 Inner Classicals, which lie just sunward of the 3:2
MMR, and 22 Centaurs, which represent an intermediate dynamical stage
between TNOs and Jupiter family comets
\citep[JFCs;][]{2008ApJ...687..714Volk}.

\section{Results}

\begin{figure} 
  \epsscale{0.9}
  \plotone{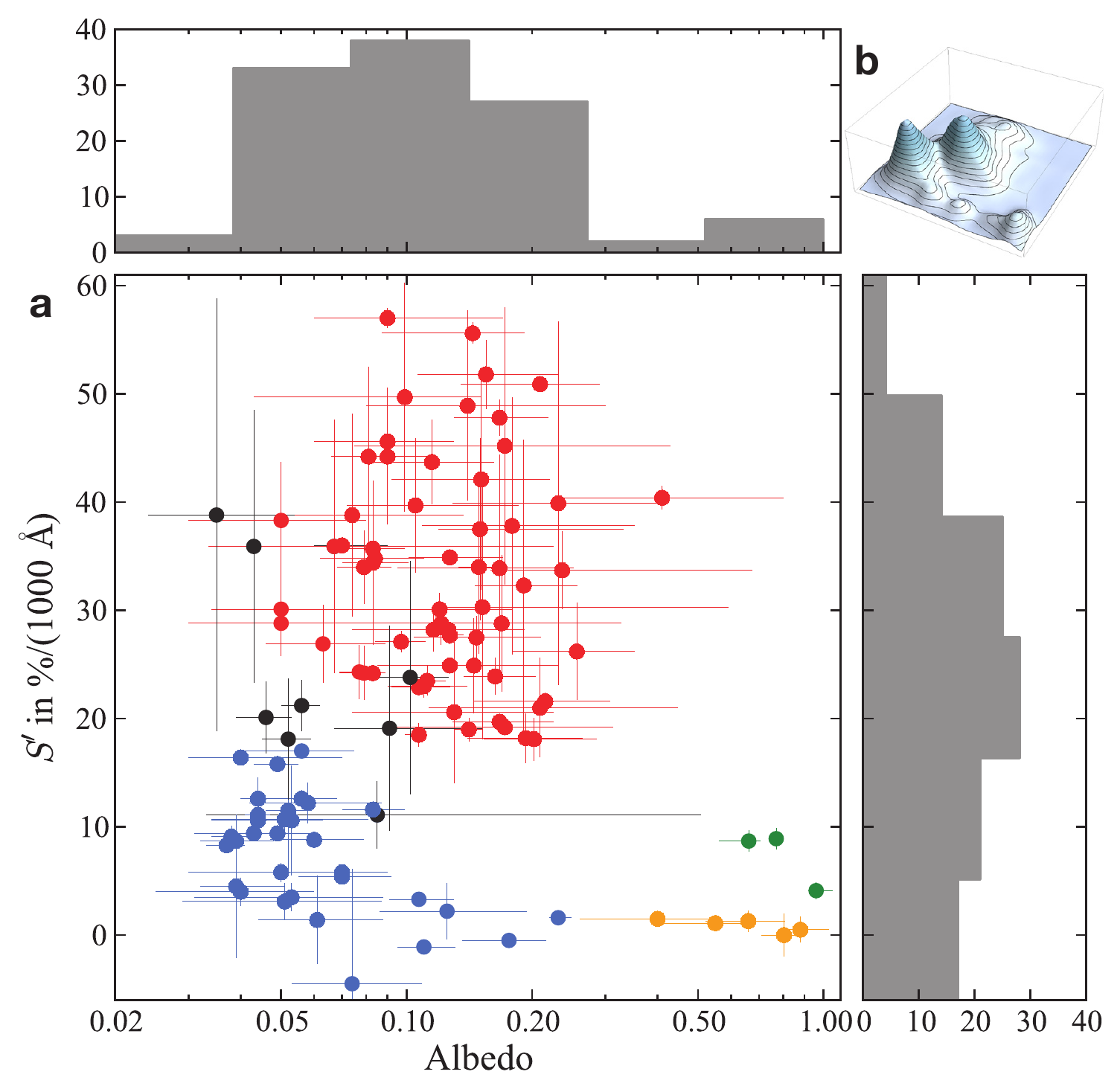}

  \caption{ a) Color-albedo diagram for 109 TNOs showing two main
    clusters, one composed of Dark Neutral objects (blue points,
    albedos $\sim 0.05$ and $S'\sim10\%$), and another of Bright Red
    objects (red points, albedo $\sim 0.15$ and $S'\sim30\%$). Black
    points have large uncertainties and ambiguous surface type.  Large
    TNOs (green) and objects with Haumea-type surfaces (orange) occupy
    a third group (bottom right). Albedo/color histograms are shown in
    gray. b) Smooth 2D histogram (Gaussian kernel, width 3\% of full
  ranges, weighted by errorbars) of the color-albedo distribution
shown in a).  } 

  \label{Fig.AS}
\end{figure}

The Herschel albedos (geometric, V-band) are shown in Figure
\ref{Fig.AS}, plotted against visible color quantified by the spectral
slope, $S'$, in units of \%/(1000~\AA) \citep{1990AJ.....99.1985Luu}.
Spectral slopes are measured directly from optical spectra when
available \citep{2009A&A...508..457Fornasier} or derived from
broadband $BR$ photometry \citep{2002AJ....123.1039Jewitt} taken from
the literature \citep{2012A&A...546A.115Hainaut,
2012A&A...546A..86Peixinho, 2013A&A...554A..49Perna}. In the remaining
text, we refer to the spectral slopes simply as ``color''. We plot
only those 109 objects for which both albedo and color are known.  The
TNOs are broadly split into two clusters: a Dark Neutral clump of
objects with low albedos and shallow spectral slopes ($p_V\sim 0.05$,
$S'\sim 10\%$), and a Bright Red agglomeration with higher albedos and
significantly redder slopes ($p_V\sim 0.15$, $S'\sim 35\%$). These two
surface types encompass $>$90\% of all midsized TNOs in our sample.  A
smaller cluster of bright neutral objects includes the dwarf planets
Eris, Pluto and Makemake which have characteristic ultra-high albedos,
plus Haumea and another 4 objects with Haumea-type surfaces. 

\begin{figure} 
  \epsscale{0.9}
  \plotone{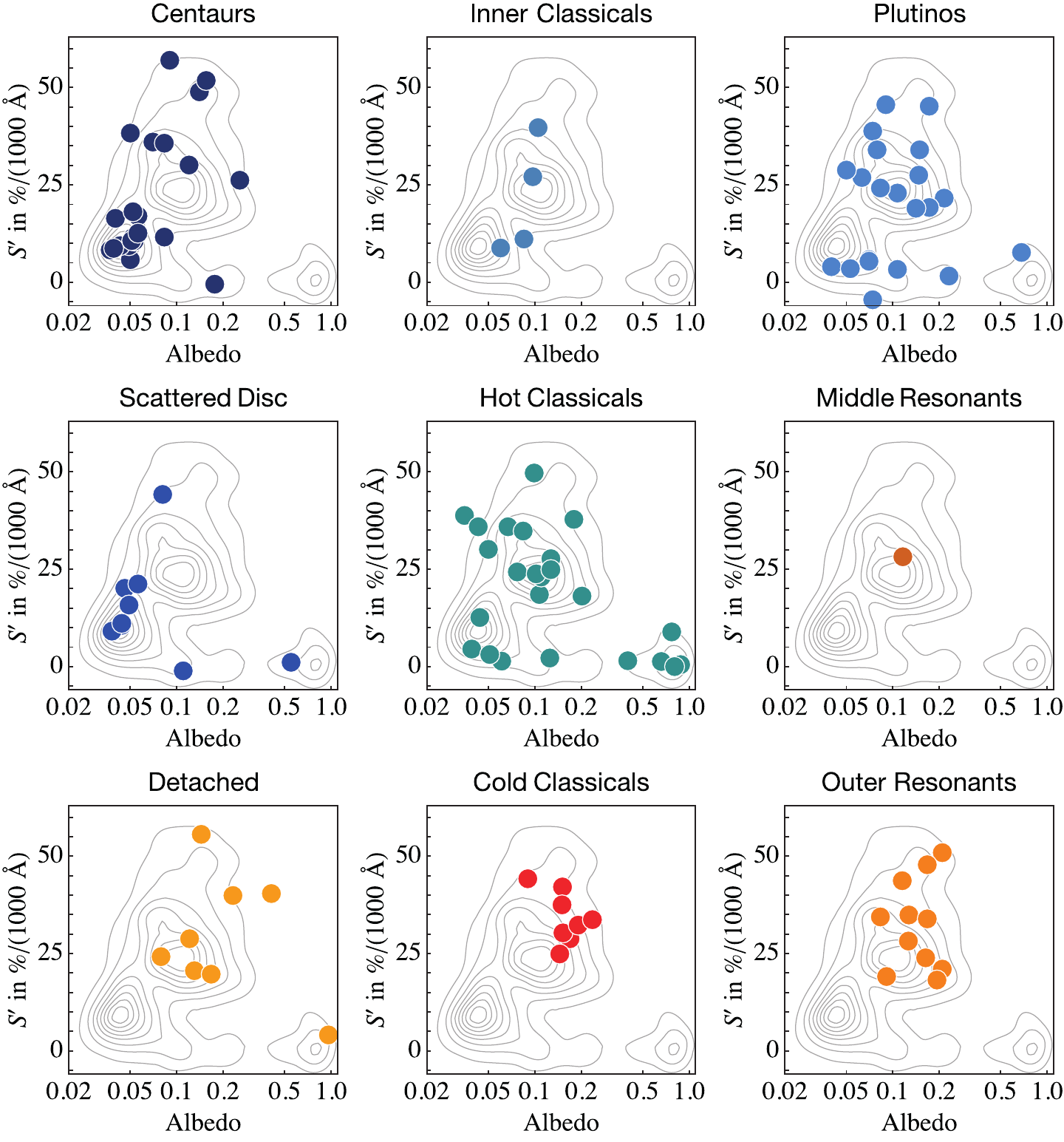}

  \caption{ Same as Figure 1 with objects gathered by dynamical class.
    Underlying contours are a density map of all objects obtained by
    bootstrap-sampling each object 100 times assuming its albedo
    (color) follows a lognormal (normal) distribution set by the
  observing uncertainties.  } 

  \label{Fig.ASGrid}
\end{figure}

The orbital distribution of Bright Red and Dark Neutral objects is
shown in Figure \ref{Fig.ae} and shows that TNOs with both surface
types exist scattered throughout the entire transneptunian region with
no obvious trend in mean heliocentric distance or closest approach to
the Sun.  We take this to indicate that the surface types are unlikely
to be set by current conditions (temperature, irradiation) that depend
on the object's distance to the Sun and are probably primitive. Figure
\ref{Fig.ASGrid} shows how TNOs in different dynamical classes are
distributed in the color-albedo diagram. Interestingly, while some
dynamical families have objects in both surface-type clusters, others
are exclusively composed of Bright Red objects.  The latter include
the Cold Classicals, the Middle and Outer Resonants, and the Detached
TNOs.  

The two main surface-type clusters in Figure \ref{Fig.AS} were
identified automatically using the {\em Mathematica 9} procedure {\tt
FindClusters[]}\footnote{http://reference.wolfram.com/mathematica/tutorial/PartitioningDataIntoClusters.html},
which we set to implement an agglomerative algorithm, using the
Euclidean distance function. Eris, Pluto, Makemake and objects with
Haumea-type surfaces were excluded leaving a set of $N=101$ TNOs.  As
seen in Figure \ref{Fig.AS}, the bulk of the TNOs falls consistently
in one of the two groups, with only a few objects near the gap between
clusters having ambiguous classification due to their large
uncertainties. To assess the significance of the two clusters we
employed three methods. Firstly, we randomly generated 1000 sets of
points in the unit square, each equal in size to the original data. We
found 2 clusters in 1.3\% of the random sets, 3 clusters in 0.1\% and
4 clusters in 0.1\% of the cases, with the remaining 98.5\% of random
sets being deemed unclustered.  Secondly, we used the gap statistic
\citep{tibshirani2001estimating} to quantify how often is a reference
distribution that better represents the data (generated along the
principal components of the data as opposed to randomly in a square)
as strongly clustered as the original. The technique selected 2
clusters as the optimal number in our data and found that the same was
true for 14 out of 1000 (1.4\%) replications of the data drawn from
the optimized reference distribution; the remaining 987 cases were
considered unclustered.  Finally, we employed the bootstrapping
technique described in \citet{efron1994introduction}.  Here, smooth
null distributions from which bootstrap replications can be drawn are
generated by convolving the normalized data with a 2D Gaussian of
width $w$.  The number of maxima of the convolved distribution
corresponds to the number of clusters in that particular bootstrap
replication. The value $w$ is selected to be the smallest that
produces a unimodal null distribution from the data. In 1000 bootstrap
replications (each with $N=101$) drawn from the null distribution, the
fraction found to have 2 or more clusters was $p=0.008$. Thus, all
three methods find a low probability ($\sim$1\%) that the clustering
seen in the data is random.

\section{Discussion}

We find that the surfaces of most TNOs fall into one of two main
types: Bright Red, or Dark Neutral (Figure \ref{Fig.AS}). Furthermore,
while some dynamical classes have both types of objects, others have
only Bright Red TNOs. The latter are those that probably originated
far from the Sun, so their characteristic surfaces would be
representative of outer solar system planetesimals.  As discussed
above, Cold Classicals likely formed in-situ and have remained
unperturbed dynamically \citep{2011ApJ...738...13Batygin}. Detached
TNOs are currently decoupled from Neptune, and some have been claimed
to be part of the inner Oort Cloud \citep{2014Natur.507..471Trujillo}.
While the origin of these objects is unknown, most scenarios suggest
that they formed beyond Neptune \citep{2002Icar..157..269Gladman}.
Middle and Outer Resonants may have been swept from nearer the Sun
during Neptune's migration.  In the classic resonance sweeping
scenario \citep{1995AJ....110..420Malhotra}, TNOs currently in the
$p$:$q$ MMR originated in the region
$a_{N}(p/q)^{2/3}<a<30(p/q)^{2/3}$, where $a_N$ is the starting point
of the Neptune's migration in AU. Assuming that Neptune began its
outward migration at $\sim$20 AU \citep{1995AJ....110..420Malhotra},
the 2:1 MMR, the innermost of the Outer Resonances, includes objects
captured at 32 AU and beyond. The alternative mechanism to populate
the MMRs invokes chaotic resonant capture during the circularisation
of Neptune's orbit following the planetary instability phase
\citep{2008Icar..196..258Levison}. An initially eccentric Neptune
would produce a chaotic sea of wide, overlapping MMRs in the
transneptunian region. As the planet's orbit circularized, the MMRs
narrowed and retained in-situ planetesimals that happened to be at
resonant locations. In summary, in the resonance sweeping scenario the
Middle and Outer Resonants are likely dominated by objects formed
beyond $\sim$30 AU, but in the chaotic capture mechanism it is
possible that they include objects formed nearer the Sun.

In contrast, Plutinos, Scattered TNOs, Inner and Hot Classicals, and
Centaurs, which are believed in most dynamic evolution models to
contain remnant planetesimals formed over a wider range of
heliocentric distances, from about 20 AU out to the current classical
Kuiper belt \citep{2011AJ....142..131Petit}, include a mixture of
Bright Red and Dark Neutral objects. This is consistent with previous
findings that different, dynamically excited TNO populations are
composed of two main types of surfaces
\citep{2012ApJ...749...33Fraser, 2013ApJ...773...22Bauer}. In the
classic sweeping resonance scenario the Plutinos were captured from
$\sim$26 AU outwards, while the Scattered Disc
\citep{2004MNRAS.355..935Morbidelli} and the Hot Classicals
\citep{2003Icar..161..404Gomes} include several objects formed at
20-30 AU and scattered following the planetary instability event and
Neptune's migration through the disk. The origin of the Inner
Classicals is uncertain, but they may be an extension of the Hot
Classical component \citep{2009AJ....137.4917Kavelaars}, while the
Centaurs probably originate in one of the populations above
\citep{2008ApJ...687..714Volk} and are expected to be physically
similar.

Previous attempts to find trends in the surface properties of TNOs
have relied mostly on broadband colors. Three key findings emerged
from the analysis of colors: that TNOs have the most diverse surfaces
of all small bodies in the solar system \citep{1996AJ....112.2310Luu},
that low eccentricity and inclination objects in the classical Kuiper
belt possess significantly redder surfaces
\citep{2000Natur.407..979Tegler} -- these objects are now recognized
as the Cold Classical population -- and that Centaurs and small,
excited TNOs are a mixture of neutrally colored and very red objects,
with no intermediate colors \citep{2003A&A...410L..29Peixinho,
2012ApJ...749...33Fraser, 2012A&A...546A..86Peixinho,
2014ApJ...782..100Fraser}.  The mixture of neutral and red objects in
these populations is taken to mean that they include planetesimals
formed at two different locations in the disk, as opposed to the
uniformly red Cold Classicals, which all formed in the distant solar
system \citep{2012ApJ...749...33Fraser}. A hypothesis to explain the
color bifurcation of objects formed at different heliocentric
distances relies on the heliocentric-distance-dependent fractionation
of surface volatiles with different sublimation temperatures
\citep{2011ApJ...739L..60Brown}. Sparser and lower quality albedo data
from Spitzer have been analyzed to reveal hints of a trend of
increasing albedo with spectral slope for Centaurs
\citep{2008prpl.conf..161Stansberry}, while a more intricate principal
component analysis of TNO albedos and colors identifies 10 different
groupings, 5 of which are red and compatible with the Bright Red type,
3 are dark and consistent with Dark Neutral surfaces, and the
remaining 2 account for the largest bodies
\citep{2013Icar..222..307DalleOre}.  Our analysis using a
significantly larger sample of high-quality albedos builds on the
trends above \citep[e.g., our albedo-color clusters closely match the
  color bifurcation first seen in][]{2012ApJ...749...33Fraser} and,
  importantly, corroborates the idea that dynamical classes believed
  to originate beyond $\sim$30 AU appear uniquely linked by a special
  combination of red color and high albedo.  

The potentially unique surfaces of TNOs formed in-situ promise to shed
light on some open questions. For instance, attempts to separate
Classical TNOs into Hot and Cold based on inclination and color
thresholds have produced inconsistent results
\citep{2008AJ....136.1837Peixinho}.  Clearly, the two components are
mixed, concealing the intrinsic inclination distribution of the
locally formed Cold population.  Careful analysis of the orbits of
Bright Red Hot Classicals may help separate the two populations in a
more robust fashion.  Another example is whether the Scattered Disk is
being replenished from other excited dynamical classes, or if it is a
fossil from the planetary instability epoch
\citep{1997Sci...276.1670Duncan}.  We find that the distribution of
Scattered TNOs in color-albedo space is unlike that of Plutinos or Hot
Classicals (significance level 98\%, 2D Kolmogorov-Smirnov test) which
naively suggests that the Scattered Disc is not being replenished from
those populations.  This is not unexpected from dynamical arguments,
but highlights the power of the surface type trends reported here.  A
detailed analysis of these questions, including the effects of object
size, is beyond the scope of this Letter and will be the subject of a
future paper. 

\begin{figure} 
  \epsscale{0.9}
  \plotone{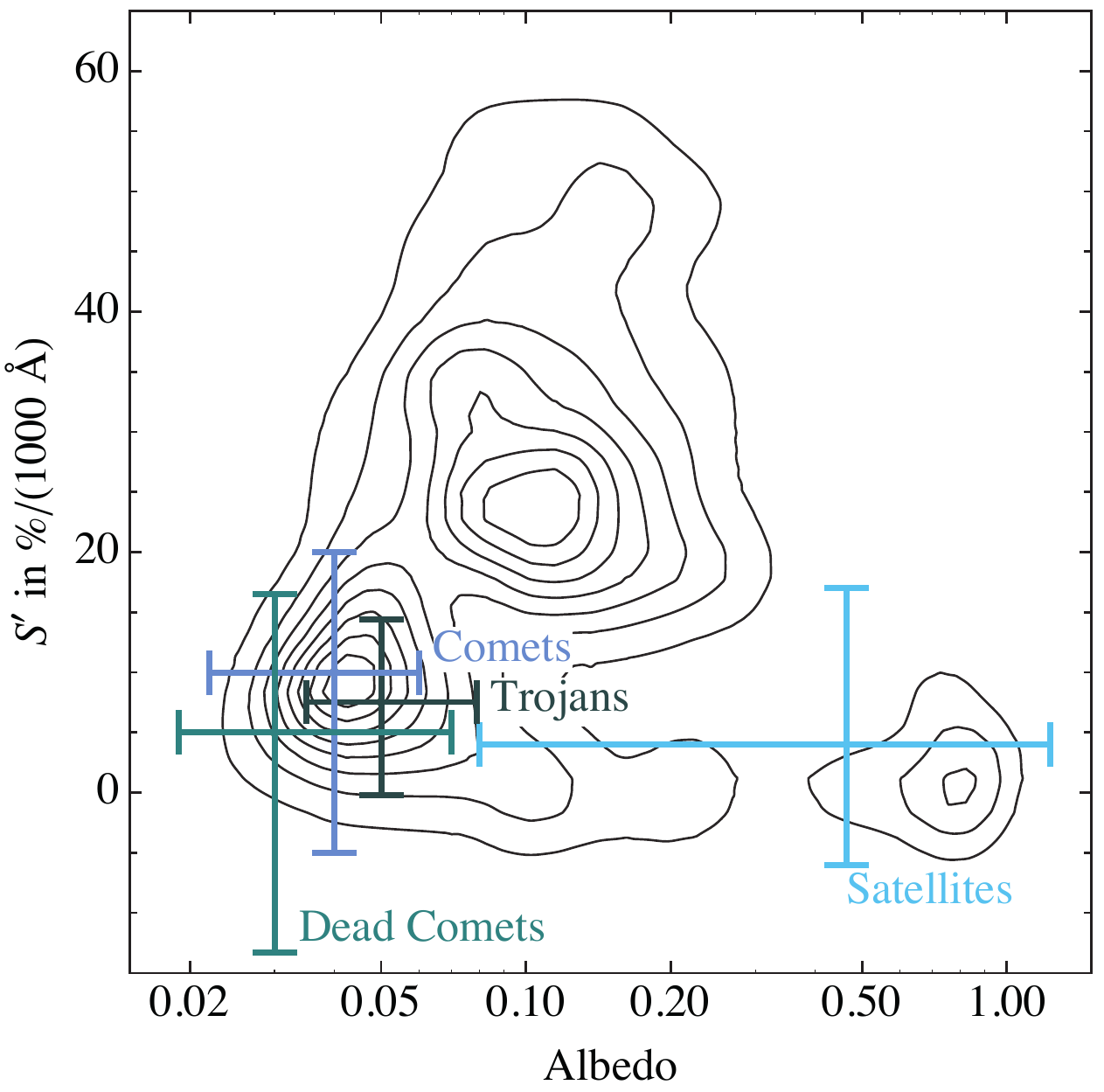}

  \caption{ Albedo and color ranges (central 95\%) for small body
    populations in the inner/intermediate solar system superimposed on
    TNO data contours (Figure \ref{Fig.ASGrid}). Comet data include
    Jupiter family and Oort cloud comets.  Jovian L4 and L5 Trojans
  are included, and satellites of the giant planets.  } 

  \label{Fig.TroComSat}
\end{figure}

No dynamical family in the outer solar system is composed solely of
Dark Neutral objects.  By contrast, objects nearer the Sun are
composed entirely of dark, neutrally colored objects (Figure
\ref{Fig.TroComSat}). JFCs, which are an end-state of the Centaurs,
and Oort cloud comets all have Dark Neutral surfaces. The same is true
for inactive objects in cometary orbits, which are believed to be dead
or dormant comets, and for Jupiter Trojans
\citep{2007Icar..190..622Fornasier}. The satellites of the giant
planets have neutrally colored surfaces, but some of the largest moons
have larger albedos.  The lack of Bright Red surfaces among inner and
intermediate solar system objects could be explained in two ways.
Either these populations contain no bodies formed in the outer solar
system, or the Bright Red material found on some TNOs is destroyed as
they approach the Sun \citep{2002AJ....123.1039Jewitt}.  The fact that
Bright Red surfaces are seen in some Centaurs but not in JFCs provides
strong evidence in favor of the latter.

We report here on evidence from albedo and color data that TNOs native
to the distant solar system ($\gtrsim$30 AU) possess unique Bright Red
surfaces, suggesting that a compositional discontinuity was in place
before the solar system was dynamically mixed. One important caveat
should be mentioned: the absence of Dark Neutral objects in our
samples of Cold Classicals, Outer Resonants and Detached TNOs may be
caused by the combination of their lower albedo (hence larger size for
the same magnitude) and the steep TNO size distribution. The current
best estimates of the TNO luminosity function suggest that the higher
albedo Bright Red objects are $\sim$3 times more likely to be included
in our sample than Dark Neutral objects
\citep{2014ApJ...782..100Fraser}.  For instance, the probability of
drawing no Dark Neutral objects in a sample of 12 Outer Resonants,
assuming they are intrinsically outnumbered 3 to 1 follows a binomial
distribution and amounts to $p=0.032$. It is thus possible that our
sampling has missed on objects that are nevertheless present. We note,
however, that the dynamical classes in our sample composed solely of
Bright Red objects taken as a whole (29 objects) lower the
aforementioned probability to $p=2.4\times10^{-4}$, beyond the typical
$3\sigma$ threshold.  Interestingly, an independent study of the
colors of resonant TNOs finds cases of neutral surfaces in the 2:1 MMR
\citep{2012AJ....144..169Sheppard}. These objects are not in our
sample and lack albedo data, but could be consistent with the Dark
Neutral group, which we do not see in the Outer Resonances. 

Our data are unable to decide whether the entire bulk composition of
planetesimals accreted at different locations varied as lower
condensation temperature volatiles became available, or only the
surface chemical makeup was modulated by the different sublimation
temperature of various volatiles. For that, we may need flyby data on
Centaurs and TNOs of both surface types and in-situ chemical sampling
of comet nuclei to come from missions such as New Horizons and
Rosetta.

\section{Conclusions}

We have analyzed the surface albedos and colors for a sample of 109
transneptunian objects. The albedo data were obtained as part of the
Herschel ``TNOs Are Cool'' survey of the outer solar system. Our main
findings are as follows.

\begin{itemize}

  \item The surfaces of most transneptunian objects fall into one of
    two types: Dark Neutral surfaces with albedos $\sim$0.05 and
    spectral slopes $S′$$\sim$10\%, and Bright Red surfaces with
    higher albedos $\sim$0.15 and significantly steeper spectral
    slopes $S′$$\sim$35\%. This clustering of surfaces lends support
    to the previously reported bifurcation in the colors of small,
    excited Kuiper belt objects and Centaurs and highlights the
    importance of albedo data for the understanding of the surface
    properties of small solar system bodies.

  \item Importantly, we find that all transneptunian objects in our
    sample thought to have formed and remained in the outer solar
    system possess Bright Red surfaces. These include the Cold
    Classical Kuiper belt objects, the Detached objects, and Resonant
    transneptunian objects in the 2:1 mean-motion resonance with
    Neptune and beyond. We note, however, that the steep
    transneptunian object size distribution may result in our small
    sample missing darker objects in these populations.

\end{itemize}

\acknowledgments PL acknowledges support from the Michael West
Fellowship and The Royal Society Newton Fellowship. Part of this work
was supported by the German \emph{DLR} project number 50 OR 1108.  RD
acknowledges support from MINECO for his Ram\'on y Cajal Contract.
CK was been supported by the Hungarian Research Fund (OTKA)
grant K-104607 and by the Bolyai Research Fellowship of the H.A.S.

Facilities: \facility{Herschel}



\vfill
\eject
\clearpage


\begin{deluxetable}{lccccc} 
  \tablehead{\colhead{Dynamical class}  & \colhead{$N$} & 
  \multicolumn{2}{c}{Albedo} & \colhead{Color} & \colhead{Surface
  Types}\\
  &     & \colhead{Median} & \colhead{CI$_{68\%}$}       &  & }
  \tablewidth{0pt}
  \tablecaption{TNO/Centaur Sample Properties}
  \startdata
  Scattered Disc   & 9   & $0.05$ & $(0.04,0.09)$ & $16.3\pm12.6$ & DN, BR \\
  Centaurs         & 22  & $0.06$ & $(0.04,0.13)$ & $21.5\pm16.5$ & DN, BR \\
  Hot Classicals   & 25  & $0.08$ & $(0.04,0.13)$ & $22.8\pm15.6$ & DN, BR \\
  Plutinos         & 20  & $0.09$ & $(0.05,0.16)$ & $20.1\pm15.4$ & DN, BR \\
  Inner Classicals & 4   & $0.09$ & $(0.06,0.18)$ & $22.4\pm12.8$ & DN, BR \\
  Middle Resonants & 1   & $0.12$ & $(0.08,0.17)$ & $28.2       $ & BR \\
  Outer Resonants  & 12  & $0.13$ & $(0.08,0.22)$ & $31.6\pm12.8$ & BR \\
  Cold Classicals  & 8   & $0.15$ & $(0.09,0.23)$ & $33.2\pm10.3$ & BR \\
  Detached TNOs    & 8   & $0.17$ & $(0.08,0.37)$ & $33.2\pm14.6$ & BR \\
  \enddata
  \tablecomments{Columns are 1) Dynamical class, 2) number of TNOs, 3)
    median albedo and 68\% confidence interval, 4) mean spectral slope
    in \%/(1000~\AA) and standard deviation. Statistics includes
    measurement uncertainties by bootstrap resampling and excludes
  dwarf planets and Haumea-type TNOs. 5) Dominant surface types present in
class (see caption for Figure~\ref{Fig.ae}).}
\label{Table.One}
\end{deluxetable}


\end{document}